\documentclass[prd,amsmath,amssymb,showpacs]{revtex4}
\begin{document}
\title{Fermion Generations from "Apple-Shaped" Extra Dimensions}

\author{Merab Gogberashvili}
\email{gogber@gmail.com}
\affiliation{Andronikashvili Institute of Physics \\
6 Tamarashvili St., Tbilisi 0177, Georgia \\and
\\ Javakhishvili State University \\
Faculty of Exact and Natural Sciences,
3 Chavchavadze Avenue, Tbilisi 0128, Georgia }

\author{Pavle Midodashvili}
\email{pmidodashvili@yahoo.com}
\affiliation{Chavchavadze State University \\
32 Chavchavadze Avenue, Tbilisi 0179, Georgia \\and
\\ Tskhinvali State University \\
2 Besiki St., Gori 1400, Georgia}

\author{Douglas Singleton}
\email{dougs@csufresno.edu}
\affiliation{Physics Dept., CSU Fresno, Fresno, CA 93740-8031,
USA}

\date{\today}

\begin{abstract}
We examine the behavior of fermions in the presence of an internal
compact 2-manifold which in one of the spherical angles exhibits a
conical character with an obtuse angle. The extra manifold can be
pictured as an apple-like surface i.e. a sphere with an extra
"wedge" insert. Such a surface has conical singularities at north
and south poles. It is shown that for this setup one can obtain,
in four dimensions, three trapped massless fermion modes which
differ from each other by having different values of angular
momentum with respect to the internal 2-manifold. The extra
angular momentum acts as the family label and these three massless
modes are interpreted as the three generations of fundamental
fermions.
\end{abstract}
\maketitle


\section{Introduction}

One of the open questions in the Standard Model of particle
physics is the fermion family puzzle - why the first generation of
quarks and leptons are replicated in two other families of
increasing mass. It is not clear how to explain the mass
hierarchy of the generations and the mixing between the families
characterized by the Cabbibo-Kobayashi-Maskawa matrix. Several
ideas have been suggested such as a horizontal family symmetry
\cite{horisontal}.

Recently the brane world idea \cite{brane} has been used to find
new solutions to old problems in particle physics and cosmology. A
key requirement for theories with extra dimensions is that the
various bulk fields (with the exception of gravity) be localized
on the brane. Brane solutions with different matter localization
mechanisms have been widely investigated in the scientific
literature \cite{local}. A pure {\it gravitational} trapping of
zero modes of all bulk fields was given in \cite{incr,Go-Si}.

The main emphasis of the present paper is to explain some
properties of fermion families in the framework of a brane model.
For the other attempts using extra dimensions see
\cite{extra-mixing,fermi}. We introduce an extra $2$-dimensional
compact surface and investigate the properties of higher
dimensional fermions place in this space-time. In $6$-dimensional
models the internal compact $2$-manifold usually is considered as
having rugby(football)-ball shaped geometry with a deficit angle
\cite{Nav,rugby}. As shown in this paper one can address the
generation puzzle using an internal $2$-surface with a profuse
angle, or having an "apple-like" geometry. Using the brane
solution of \cite{Go-Si} we show that for apple-shaped extra
dimensions three fermion generations naturally arise from the zero
modes of a {\it single} $6$-dimensional spinor field. This gives a
purely geometrical mechanism for the origin of three generations
of the Standard Model fermions from one generation in a
higher-dimensional theory. The localized fermions are stuck at
different points in the extra space similar to the model
\cite{fermi}. A mass hierarchy and mixings between the three zero
modes are obtained by introducing of a Yukawa-type coupling to a
single $6$-dimensional scalar field.


\section{Solution of $6$-dimensional Einstein equations}

In this article we consider $6$-dimensional space-time with the
signature $(+ - - - - -)$. Einstein's equations in this space have
the form
\begin{equation}\label{EinsteinEquation}
R_{AB} - \frac{1}{2}g_{AB} R = \frac{1}{M^4}\left( g_{AB} \Lambda
+ T_{AB} \right)~,
\end{equation}
where $M$ and $\Lambda$ are the $6$-dimensional fundamental scale
and the cosmological constant. Capital Latin indices run over $A,
B,... = 0, 1, 2, 3, 5, 6 $.

To split the space-time into $4$-dimensional and $2$-dimensional
parts we use the metric {\it ansatz}
\begin{equation}\label{MetricAnsatz}
ds^2  = \phi ^2 \left( \theta  \right)g_{\mu \nu } \left(
{x^\alpha  } \right)dx^\mu  dx^\nu   - \varepsilon ^2 \left(
{d\theta ^2  + b^2\sin ^2 \theta d\varphi ^2 } \right)~,
\end{equation}
where $\varepsilon$ and $b$ are constants. Here the metric of
ordinary $4$-space-time, $g_{\mu \nu } \left( {x^\alpha  }
\right)$, has the signature $(+ - - -)$ (the Greek indices
$\alpha, \mu, \nu ... =0,1,2,3$ refer to $4$-dimensional
coordinates). The extra compact $2$-manifold is parameterized by
the two spherical angles $\theta$ and $\varphi$ ($0 \le \theta \le
\pi ,\,\,0 \le \varphi \le 2\pi$). We take this $2$-surface to be
attached to the brane at the point $\theta = 0$. Thus the geodesic
distance into the extra dimensions goes from north to south pole
of the extra $2$-spheroid when $\theta$ changes from $0$ to $\pi$.

If in \eqref{MetricAnsatz} the constant $b = 1$ then the extra
$2$-surface is exactly a $2$-sphere with the radius $\varepsilon$.
If $b \ne 1$ the extra manifold is a $2$-spheroid with either a
deficit or profuse angle $\varphi$, i.e. its conical sections,
$\theta = const$, are either missing some angle, $\delta \varphi$,
or have some extra angle, $\delta \varphi$. The metric for this
$2$-manifold will take usual form with $b = 1$ if we redefine
$\varphi$ so it ranges from $0$ to $2\pi b$. One can think of the
extra $2$-surface as being of sphere with cut out (if $b < 1$), or
inserted (if $b > 1$) the "wedge" having an angle $\delta \varphi
= 2 \pi (b-1)$. This gives a $\delta$-like contribution to the
curvature tensor localized at the points with $\sin \theta = 0$.
These singularities can be canceled by introduction of $3$-branes
at these positions \cite{Nav}. Usually in the literature one
considers the case $b < 1$ with the deficit angle leading to
rugby(football)-ball shaped geometry \cite{rugby}. As it will be
clear below we need the case $b>1$ which gives a profuse angle.
Thus the extra $2$-manifold can be imagined as the apple-like
surface.

The {\it ansatz} for the energy-momentum tensor of the bulk matter
fields we take in the form
\begin{equation}\label{BulkMatterAnsatz}
T_{\mu \nu }  =  - g_{\mu \nu } E\left( \theta  \right), ~~~~~T_{ij} =
- g_{ij} P\left( \theta  \right), ~~~~~ T_{i\mu }  = 0~.
\end{equation}
small Latin indices correspond to the two extra coordinates. The
source functions $E$ and $P$ depend only on the extra coordinate
$\theta$.

For these {\it ans{\"a}tze} Einstein's equations
(\ref{EinsteinEquation}) take the following form:
\begin{eqnarray}\label{Einstein'sEquations}
3\frac{\phi ''}{\phi} + 3\frac{\phi '^2 }{\phi ^2 } + 3\frac{\phi
'}{\phi }\cot\theta - 1 =\frac{\varepsilon^2}{M^4}
\left[ E\left(\theta\right) - \Lambda \right], \nonumber\\
6\frac{\phi '^2 }{\phi ^2} + 4\frac{\phi '}{\phi}\cot\theta =
\frac{\varepsilon ^2}{M^4}\left[ P\left(\theta\right)-\Lambda\right],\\
4\frac{\phi ''}{\phi} + 6\frac{\phi '^2}{\phi ^2} =
\frac{\varepsilon^2}{M^4}\left[ P\left(\theta\right)-\Lambda
\right], \nonumber
\end{eqnarray}
where the prime denotes differentiation $d/d\theta$. For the
$4$-dimensional space-time we have assumed zero cosmological
constant and Einstein's equations in the form
\begin{equation}
R_{\alpha \beta }^{\left( 4 \right)}  - \frac{1}{2}g_{\alpha \beta
} R^{\left( 4 \right)}  = 0,
\end{equation}
where $R_{\alpha \beta }^{\left( 4 \right)}$ and $R^{\left( 4
\right)}$ are $4$-dimensional Ricci tensor and scalar curvature.

In \cite{Go-Si} a non-singular solution of
(\ref{Einstein'sEquations}) was found for the boundary conditions
$\phi (0) =1, ~ \phi '(0) = 0$. The solution was given by
\begin{equation}
\phi \left( \theta  \right) = 1 + \left( {a - 1} \right)\sin ^2
(\theta /2 )~,
\end{equation}
where $a$ is the integration constant. The source terms for this
solution were given by
\begin{equation}
E (\theta) = \Lambda \left[ \frac{3(a+1)}{5 \phi ( \theta )} -
\frac{3 a}{10 \phi ^2(\theta )} \right] ~, ~~~ P (\theta) =
\Lambda \left[ \frac{4(a+1)}{5 \phi ( \theta )} - \frac{3 a}{5
\phi ^2(\theta )} \right]
\end{equation}
and with the radius of the extra $2$-spheroid given by
$\varepsilon^2 = 10M^4 /\Lambda$. For simplicity in this paper we
take $a = 0$ so that below we will use the warp factor
\begin{equation}\label{Solution}
\phi \left( \theta  \right) = 1 - \sin ^2 (\theta /2 ) = \cos ^2
(\theta /2) ~.
\end{equation}
This warp factor equals one at the brane location ($\theta = 0$)
and decreases to zero in the asymptotic region $\theta = \pi$,
i.e. at the south pole of the extra $2$-dimensional spheroid.

The expression for the determinant of our {\it ansatz}
(\ref{MetricAnsatz}), which will be used often in what follows, is
given by
\begin{equation} \label{determinant}
\sqrt{-g} = \sqrt{-g^{(4)}} \varepsilon^2 \phi^4 (\theta) \sin \theta~,
\end{equation}
where $\sqrt{-g^{(4)}} $ is determinant of $4$-dimensional
space-time.


\section{Fermions in six dimension}

Let us consider spinors in the $6$-dimensional space-time
(\ref{MetricAnsatz}), where the warp factor $\phi (\theta)$ has
the form \eqref{Solution}. The action integral for the
$6$-dimensional massless fermions in a curved background is
\begin{equation}\label{FermionAction}
S_{\Psi} = \int d^6 x \sqrt { - g} \left[ i\overline \Psi
h_{\widetilde A}^B\Gamma ^{\widetilde A} D_B \Psi  + h. c. \right] ~,
\end{equation}
$D_A$ denote covariant derivatives, $\Gamma ^{\widetilde A}$ are
the $6$-dimensional flat gamma matrices and we have introduced the
{\it sechsbein} $h_A^{\widetilde A}$ through the usual definition
\begin{equation}
g_{AB} = h_A^{~\widetilde A} h_B^{~\widetilde B} \eta _{\widetilde
A\widetilde B}~,
\end{equation}
$\widetilde A,\widetilde B,...$ are local Lorentz indices.

In six dimensions a spinor
\begin{equation}\label{Psi8}
\Psi (x^A) = \left(\begin{array}{ll}
\psi \\ \xi
\end{array}\right)
\end{equation}
has eight components and is equivalent to a pair of
$4$-dimensional Dirac spinors, $\psi$ and $\xi$.

In this paper we use the following representation of the flat
$(8\times 8)$ gamma-matrices (for simplicity we drop the tildes on
the local Lorentz indices when no confusion will occur)
\begin{equation} \label{Gamma}
\Gamma_\nu=\left(\begin{array}{cc}
\gamma_\nu & 0 \\
0 & -\gamma_\nu
\end{array}\right)~, ~~~~~ \Gamma_\theta
=\left(\begin{array}{cc}
0 & -1 \\
1 & 0
\end{array}\right)~, ~~~~~\Gamma_\varphi = \left(\begin{array}{cc}
0 & i \\
i & 0
\end{array}\right)~,
\end{equation}
where $1$ denotes the $4$-dimensional unit matrix and $\gamma_\nu$
are ordinary $(4 \times 4)$ gamma-matrices. It is easy to check
that the representation (\ref{Gamma}) gives the correct space-time
signature $(+-----)$. The $6$-dimensional analog of the $\gamma_5$
matrix in the representation (\ref{Gamma}) has the form
\begin{equation} \label{Gamma7}
\Gamma_7=\left(\begin{array}{cc}
\gamma_5 & 0 \\
0 & \gamma_5
\end{array}\right)~.
\end{equation}
From (\ref{Psi8}) one finds that the $6$-dimensional left-handed
(right-handed) particles correspond to a pair of $4$-dimensional
particles, $\psi$ and $\xi$ which correspond to the particle
(anti-particle) from the $4$-dimensional point of view.

The $6$-dimensional massless Dirac equation, which follows from
the action (\ref{FermionAction}), has the form
\begin{equation} \label{dirac6}
\left(h^\mu _{\widetilde {B}} \Gamma ^{\widetilde B} D_{\mu} +
h^\theta _{\widetilde {B}} \Gamma ^{\widetilde B} D_\theta +
h^\varphi _{\widetilde {B}} \Gamma ^{\widetilde B} D_{\varphi}
\right) \Psi (x^A)= 0 ~,
\end{equation}
with the {\it sechsbein} for our background metric
\eqref{MetricAnsatz} given by
\begin{equation} \label{h}
h_{\widetilde A}^{~B}=\left( \frac{1}{\phi} \delta_{\widetilde
\mu}^B,~ \frac{1}{\varepsilon} \delta_{\widetilde \theta}^B
,~ \frac{1}{b\varepsilon \sin \theta} \delta_{\widetilde \varphi}^B
\right)~.
\end{equation}
From the definition
\begin{equation} \label{spin1}
\omega^{ \widetilde M\widetilde N}_M = \frac{1}{2} h^{N\widetilde
M} \left(\partial_M h^{\widetilde N}_N -\partial_N h^{\widetilde
N}_M \right) - \frac{1}{2} h^{N\widetilde N}\left(\partial_M
h^{\widetilde M}_N - \partial_N h^{\widetilde M}_M\right)-
\frac{1}{2} h^{P\widetilde M}h^{Q\widetilde N}\left(
\partial_P h_{Q\widetilde R} -
\partial_Q h_{P\widetilde R}\right) h^{\widetilde R}_M
\end{equation}
the non-vanishing components of the spin-connection for the {\it
sechsbein} (\ref{h}) can be found
\begin{equation}
\omega _\varphi ^{\widetilde \theta\widetilde\varphi } = - b\cos
\theta, ~~~~~ \omega_\nu^{\widetilde \theta\widetilde\nu} = -
\frac{\phi'}{\varepsilon} = \frac{\sin \theta}{ 2\varepsilon}~.
\end{equation}
The covariant derivatives of the spinor field have the forms
\begin{eqnarray} \label{covariant}
D_\mu \Psi (x^A) = \left[ \partial_\mu + \frac{\sin \theta
}{4\varepsilon} \Gamma _\theta\Gamma _\nu
\right] \Psi (x^A)~ , \nonumber \\
D_\theta \Psi (x^A) = \partial_\theta \Psi (x^A) ~ ,\\
D_\varphi \Psi (x^A) = \left(\partial_\varphi - \frac{b\cos
\theta}{2} \Gamma _\theta\Gamma _\varphi \right) \Psi (x^A)
~.\nonumber
\end{eqnarray}

Then Dirac's equation \eqref{dirac6} takes the form
\cite{Ca-Hi,Abr}
\begin{eqnarray} \label{dirac}
\left[ \frac{1}{\phi}\Gamma^\mu \frac{\partial}{\partial x_\mu}
+\frac{\sin \theta}{4\varepsilon\phi} \Gamma^\nu\Gamma
_\theta\Gamma _\nu + \frac{1}{\varepsilon}\Gamma^\theta
\frac{\partial}{\partial\theta} + \frac{1}{b\varepsilon \sin
\theta}\Gamma^\varphi \frac{\partial}{\partial\varphi} -
\frac{\cot \theta} {2\varepsilon}\Gamma^\varphi
\Gamma_\theta\Gamma_\varphi \right]\Psi (x^A) = \nonumber
\\
= \left[ \frac{1}{\phi}\Gamma^\mu \frac{\partial}{\partial x_\mu}
+ \frac{1}{\varepsilon} \Gamma^\theta \left(
\frac{\partial}{\partial\theta} - \frac{\sin \theta}{\phi} +
\frac{\cot \theta}{2} \right) + \frac{1}{b\varepsilon \sin \theta}
\Gamma^\varphi \frac{\partial}{\partial\varphi} \right] \Psi (x^A)
= 0 ~.
\end{eqnarray}

This system of first order partial differential equations can be
treated using the following separation of variables
\begin{equation}\label{Psi}
\Psi (x^A)= \sum_{l}
\frac{e^{il\varphi}}{\sqrt{2\pi} \phi^2(\theta)}
\left(\begin{array}{ll}\alpha_l(\theta)\psi_l (x^\nu)\\
\beta_l(\theta)\xi_l (x^\nu)\end{array}\right),
\end{equation}
where $\psi_l (x^\nu)$ and $\xi_l (x^\nu)$ are $4$-dimensional
Dirac spinors. Here we note that since dimension of $\Psi (x^A)$
in six dimensions is $m^{5/2}$ then dimensions of
$\alpha_l(\theta),~ \beta_l(\theta)$ and $\psi_l (x^\nu),~ \xi_l
(x^\nu)$ should be $m$ and $m^{3/2}$ respectively.

At the end of the section we note that our case is unlike the
model studied in \cite{Abr}, which examined spin-$1/2$ particles
confined on a $2$-sphere. In our case the internal $2$-manifold is
only a part of the bulk $6$-dimensional space-time and we are
looking for spinors in four dimensions. It is the functions
$\psi_l (x^\nu)$ and $\xi_l (x^\nu)$ in (\ref{Psi}) which must
have spinor representations. So the wave-function given in
(\ref{Psi}) is single-valued for $2\pi$ rotations around the brane
by the extra angle $\varphi$. Thus the quantum number $l$ takes
integer values -- $l = 0, \pm 1, \pm 2, ... $ -- and not
half-integer values.


\section{Fermion generations}

We are looking for $4$-dimensional fermionic zero modes. To this
end we take $\psi_l(x^\nu)$ and $\xi_l (x^\nu)$ in (\ref{Psi}) to
obey the $4$-dimensional, massless Dirac equations
\begin{equation} \label{dirac-brane}
\gamma^\mu \partial_\mu \psi_l (x^\nu) = \gamma^\mu \partial_\mu
\xi_l (x^\nu) = 0~.
\end{equation}
There will also be very massive KK modes whose masses will go a
integer multiples of the inverse size of the extra $2$-dimensional
space i.e. as $1/\varepsilon $. However, we will assume later that
$1/\varepsilon \simeq 1$ TeV. Thus these massive KK modes have a
much higher mass and are distinct from the three fermion
generations. For the massless case the $4$-spinors $\psi_l(x^\nu)$
and $\xi_l (x^\nu)$ are indistinguishable from the $4$-dimensional
point of view and we can write $\psi_l(x^\nu) =  \xi_l (x^\nu)~$

Inserting (\ref{Psi}) and (\ref{dirac-brane}) into (\ref{dirac})
converts the bulk Dirac equation into
\begin{equation} \label{dirac2}
\left[  \Gamma^\theta \left( \frac{\partial}{\partial\theta} +
\frac{\cot \theta}{2} \right) + \frac{il}{b\sin \theta}
\Gamma^\varphi \right]
\left(\begin{array}{ll}\alpha_l(\theta) \\
\beta_l(\theta) \end{array}\right) = 0 ~.
\end{equation}
Using the representation for $\Gamma ^\theta$, $\Gamma ^\varphi$
gives the following system of equations for $\alpha_l(\theta)$ and
$\beta_l(\theta)$
\begin{eqnarray} \label{dirac3}
\left( \frac{\partial}{\partial\theta} + \frac{\cot \theta}{2} -
\frac{l}{b\sin \theta} \right) \alpha_l(\theta) = 0~, \nonumber \\
\left( \frac{\partial}{\partial\theta} + \frac{\cot \theta}{2} +
\frac{l}{b\sin \theta} \right) \beta_l(\theta)  = 0 ~.
\end{eqnarray}
The solutions of these equations are
\begin{equation}\label{a-b}
\alpha_{l}(\theta) = A_l\frac{\tan^{l/b}(\theta /2)}{\sqrt{\sin
\theta}} ~, ~~~~~ \beta_{l}(\theta) = B_l\frac{\tan^{-l/b}(\theta
/2)}{\sqrt{\sin\theta}}~,
\end{equation}
where $A_l$ and $B_l$ are integration constants with the dimension
of mass.

The normalizable modes are those for which
\begin{equation}
\label{normalize} \int \sqrt{-g} ~ d^6 x ~ {\bar \Psi} \Psi = \int
\sqrt {g^{(4)}} ~ d^4 x ~ \left( {\bar \psi _l}{\psi _l} + {\bar
\xi _l}{\xi _l} \right) ~.
\end{equation}
In other words we want the integral over the extra coordinates,
$\varphi$ and $\theta$, to equal $1$. Thus inserting (\ref{Psi}),
(\ref{a-b}) and the determinant (\ref{determinant}) into
(\ref{normalize}) the requirement that the integral over $\varphi$
and $\theta$ equal $1$ gives
\begin{equation}\label{normalization}
\varepsilon^2 \int_0^\pi d\theta \left[ A_l^*A_l \tan^{2l/b}
(\theta /2) + B_l^*B_l \tan^{-2l/b} (\theta /2) \right] = 1 ~,
\end{equation}
where the integral over $\varphi$ contributes $2\pi$.

Using the formula
\begin{equation}\label{int}
\int_0^\pi d\theta ~\tan^{2c}(\theta /2) = \pi/\cos (c\pi)~,
~~~~~~~~ -1 < 2c <  1
\end{equation}
we see that (\ref{normalization}) is convergent only for the case
\begin{equation} \label{b}
-b < 2l < b~.
\end{equation}
Recall that the parameter $b$ in (\ref{MetricAnsatz}) is an
integration constant of Einstein's equations and governs the
topology of the internal $2$-spheroid. If $b = 1$ the internal
$2$-surface is exactly a sphere. For this case, as it clear from
(\ref{b}), there exist only one zero mode with $l = 0$. If on the
other hand $2 < b \leq 4 $ we have exactly three fermionic zero
modes with the quantum numbers $l = 0$ and $l = \pm 1$. To be
concrete we will set $b = 4$ in the following. Other choices of
$b$ from this interval will only slightly change the numerical
results below. From the normalization condition
(\ref{normalization}) we now find the following relation for the
constants $A_l$ and $B_l$
\begin{equation}\label{norm}
\pi\varepsilon^2 (A^*_lA_l + B^*_lB_l)  = \cos (l\pi /4)~,
\end{equation}
where $l= 0, \pm 1$.

Explicitly the expressions for the three normalizable $8$-spinors
(\ref{Psi}) that solve the $6$-dimensional Dirac equations
(\ref{dirac}) are
\begin{eqnarray}\label{3-Psi}
\Psi_{0} (x^A)= \frac{1}{\sqrt{2\pi \sin\theta}~\phi^2(\theta)}
\left(\begin{array}{ll} A_0
\\
B_0
\end{array}\right)\psi_{0} (x^\nu)~, \nonumber \\
\Psi_{1} (x^A)= \frac{1}{\sqrt{2\pi
\sin\theta}~\phi^2(\theta)}~e^{i\varphi } \left(\begin{array}{ll}
\tan^{1/4}(\theta/2)A_1
\\
 \tan^{-1/4}(\theta/2)B_1
\end{array}\right)\psi_{1} (x^\nu)~, \\
\Psi_{-1} (x^A)= \frac{1}{\sqrt{2\pi
\sin\theta}~\phi^2(\theta)}~e^{-i\varphi } \left(\begin{array}{ll}
\tan^{-1/4}(\theta/2)A_{-1}
\\
 \tan^{1/4}(\theta/2)B_{-1}
\end{array}\right)\psi_{-1} (x^\nu)~,\nonumber
\end{eqnarray}
where the constants $A_l$ and $B_l$ obey the relations
(\ref{norm}).

These three normalizable modes all appear as massless
$4$-dimensional fermions on the brane. To explain the observed
mass spectrum and mixing between these fermions one needs to
couple these particles to a scalar (Higgs) field.


\section{Coupling with Higgs field}

In the previous section it was shown that by adjusting the
integration constant $b$ in our gravitational background
(\ref{MetricAnsatz}) it is possible to get three zero-mass modes
on the brane. To make this model more realistic we have two
problems:

a) There is no mixing between the different generations due to the
orthogonality of the angular parts of the higher dimensional wave
functions. Overlap integrals like $\int d\varphi~{\bar \psi _l}
\psi_{l'}$, which characterize the mixing between the different
states, vanish since
\begin{equation} \label{Int}
\int _0 ^{2 \pi} d\varphi~ e^{-il\varphi} e^{il'\varphi} = 0 ~.
~~~~~ l \ne l'
\end{equation}

b) All the fermionic states (\ref{3-Psi}) are massless, whereas
the fermions of the real world have masses that increase with each
family.

Following \cite{neronov} we address both of these issues by
introducing a coupling between the fermions with the bulk scalar
field $\Phi_p (x^A)$ (which has dimensions (mass)$^2$) by adding to the
action an interaction term of the form
\begin{equation}
\label{S-int}
S_{int} = \frac 1F \int d^4 x d\varphi d\theta
\sqrt{- g} ~\Phi_p \bar {\Psi} _l \Psi _{l'} ~,
\end{equation}
$F$ is the coupling constant between the scalar and spinor
fields and has the dimensions of mass.

For simplicity we take the massless, real scalar field to be of the
form
\begin{equation} \label{higgs}
\Phi_p (x ^A) = \kappa_p ~\Phi_p (\theta)~ e^{ip \varphi}~,
\end{equation}
i.e. the scalar field only depends on the bulk coordinates
$\theta, \varphi$, not on the brane coordinates $x ^\mu$. In
(\ref{higgs}) the angular quantum number $p$ is an integer and
$\kappa_p$ are the $4$-dimensional constant parts of $\Phi_p (x ^A)$
having dimensions of mass.

The equation of motion of a massless real scalar field in six
dimensions has the form:
\begin{equation} \label{scalar}
\frac{1}{\sqrt{-g}} D_A \left[\sqrt{-g}~ g^{AB} ~D_B \Phi (x
^A)\right] = 0~.
\end{equation}
Using the form of Laplace operator on our $2$-spheroid
\begin{equation}\label{laplace2}
\Delta_2 = - \frac{1}{\varepsilon^2}\left(
\frac{\partial^2}{\partial\theta^2} + \cot \theta
\frac{\partial}{\partial\theta} + \frac{4
\phi^\prime}{\phi}\frac{\partial}{\partial\theta}+
\frac{1}{b^2\sin^2\theta} \frac{\partial^2}{\partial\varphi^2}
\right)~,
\end{equation}
where $\phi$ is given by (\ref{Solution}), the equation
(\ref{scalar}) can be written as
\begin{equation}
\Phi_p ^{\prime \prime} + \left( \cot \theta - \frac{4 \sin \theta
}{1 + \cos \theta } \right) \Phi_p ^\prime -\frac{p^2}{b^2\sin^2
\theta}\Phi_p = 0~.
\end{equation}
It is possible to give an exact solution to this equation in
quadratures. However, this solution is a complicated function. In
order to make understandable estimates of the masses and mixings
we will use approximate solutions. Close to the origin ($\theta
\rightarrow 0$), when $\sin \theta \rightarrow 0$ and $\phi
\rightarrow 1 $ this equation can be approximated as
\begin{equation} \label{Phi'}
\Phi_p ^{\prime \prime} +  \cot \theta \Phi_p ^\prime -
\frac{p^2}{b^2\sin^2 \theta } \Phi_p = 0~.
\end{equation}
For $p = 0$ a solution to this equation is
\begin{equation} \label{Phi-0}
\Phi_0 (\theta) = D_0 \left\{ 1+ \ln \left[ \tan (\theta/2)\right]
\right\} ~, ~~~~~ p = 0
\end{equation}
where $D_0$ is an integration constant.

For non-zero $p$ one of the solutions of \eqref{Phi'} is
\begin{equation} \label{Phi-p}
\Phi_p (\theta) = D_p \cosh \left\{ \frac{p}{b}\ln \left[ \cot
(\theta/2)\right]\right\} ~, ~~~~~ p \ne 0
\end{equation}
where $D_p$ are integration constants. Note that these solutions
(as well as the spinor fields (\ref{Psi}) and (\ref{a-b})) are
singular at $\sin \theta = 0$, however, because of the determinant
(\ref{determinant}) the various integrals done with these fields
are finite.

We determine the constants $D_p$ by requiring that the scalar
field is normalized over the extra coordinates, i.e. using
(\ref{determinant}) we require
\begin{equation} \label{Phi-norm}
2\pi \varepsilon^2 \int_0^\pi d\theta ~\sin \theta ~\phi^4
(\theta) ~\Phi_p^2 (\theta)= 1~.
\end{equation}

For the values of $a$ and $b$ used in this paper ($a=0$, $b=4$)
from (\ref{Phi-norm}) we find
\begin{equation}\label{D}
D_0 = \frac{1}{\varepsilon\sqrt{\frac{\pi^3}{15} - \frac{17
\pi}{60}}} \approx \frac{0.92}{\varepsilon}~, ~~~~~ D_{\pm 1} =
\frac{1}{\varepsilon \sqrt{\frac{2 \pi}{5}+ \frac{447
\sqrt{2}\pi^2}{4096}}} \approx \frac{0.60}{\varepsilon}~, ~~~~~
D_{\pm2} = \frac{1}{\varepsilon \sqrt{\frac{2 \pi}{5}+ \frac{35
\pi^2}{128}}} \approx \frac{0.50}{\varepsilon}~.
\end{equation}

Substituting (\ref{higgs}) and (\ref{Psi}) into (\ref{S-int}) we
find
\begin{equation}
\label{S-inter} S_{int} = U_{l, l'}^p \int d^4 x ~\sqrt{-
g^{(4)}}~\bar{ \psi} _l (x ^\mu ) \psi _{l'} (x ^\mu)~,
\end{equation}
with
\begin{equation} \label{U}
U_{l,l'}^p = \frac{\varepsilon^2 \kappa_p}{2\pi F} \int_0^{2\pi}
d\varphi e^{i (p-l+l') \varphi} \int_0^\pi d \theta \sin \theta
\Phi_p (\theta) \left[ A_l^*A_{l'}\alpha_l (\theta) \alpha_{l'}
(\theta)+ B_l^*B_{l'}\beta_l (\theta)\beta_{l'} (\theta)\right] ~,
\end{equation}
where $D_p$ are expressed by (\ref{D}) and $A_l$, $B_l$ obey the
relations (\ref{norm}). Below we will use the new definition $f_p
= \kappa_p /F$ for the ratios of the $4$-dimensional constant
values of Higgs field from (\ref{higgs}) and of the coupling
constant from (\ref{Int}).

The first integral in (\ref{U}) for the quantities $U_{ll'}^p$
will be non-zero if
\begin{equation} \label{p-l}
p-l+l' = 0 ~.
\end{equation}
When $l=l'$ and $p=0$ this gives a mass term; when $l \ne l'$ and
$p \ne 0$ this gives mixings between the $l$ and $l'$ modes.


\section{Masses and mixings}

To find mass terms appearing because of coupling of the three
fermionic zero modes (\ref{3-Psi}) with the Higgs field
(\ref{higgs}) for the angular momentum quantum numbers in
(\ref{U}) we should use the values, $p = 0~,~ l = l'$, or
calculate only the components of the matrix (\ref{U}) with the
zero upper index. Using (\ref{a-b}) and (\ref{Phi-0}) from
(\ref{U}) we get
\begin{eqnarray}
\label{U0} U_{0,0}^0&=&f_0D_0\varepsilon^2\pi\left(
A_0^*A_0+B_0^*B_0\right)=
f_0 D_0~,\nonumber\\
U_{1,1}^0&=&f_0\frac{D_0\varepsilon^2 \pi}{\sqrt{2}} \left[ (2 +
\pi)A_1^*A_1 + (2 - \pi )B_1^*B_1\right] = f_0 D_0 \left(
\frac{2-\pi}{2} + \sqrt{2}\varepsilon ^2\pi ^2|A _1|^2\right)~,\\
U_{-1,-1}^0&=&f_0\frac{D_0 \varepsilon^2 \pi}{\sqrt{2}} \left[ (2
- \pi)A_{-1}^*A_{-1}+(2+\pi )B_{-1}^*B_{-1}\right]= f_0D_0\left(
\frac{2+\pi}{2}-\sqrt{2}\varepsilon^2\pi^2|A _{-1}|^2\right)~.
\nonumber
\end{eqnarray}
To obtain the last equality in each term above we have used
(\ref{norm}) to eliminate $|B_{\pm 1}|^2$ in favor of $|A_{\pm
1}|^2$.

As a concrete example of how the realistic mass hierarchy can
arise let us take $1 / \varepsilon \simeq 1$ TeV so that from
(\ref{D}) we have $D_0 \simeq 1$ TeV. This choice is made so that
the massive KK modes (whose mass $\simeq 1/\varepsilon$) will be
much heavier than the three zero mass modes, even after they are
given a mass via the Higgs mechanism. Next let us examine three
quarks from the "down" sector, i.e. $d$, $s$ and $b$ quarks. This
is meant as a toy model since we do not have an "up" sector and we
do not have three generations of leptons. Our aim here is just to
show that it is possible to generate a realistic fermion mass
hierarchy from an extra dimensional model.

Making the association that $s$-quark $\rightarrow ~ l=-1$,
$b$-quark $\rightarrow ~ l=0$ and  $d$-quark $\rightarrow ~ l=+1$,
we get the following conditions on $U_{l,l}^0$ from (\ref{U0})
\begin{equation} \label{mass}
U_{-1,-1}^0 = m_s \approx 100~MeV~, ~~~~~ U_{0,0}^0 = m_b \approx
4200 ~MeV~, ~~~~~U_{1,1}^0 = m_d \approx 5 ~MeV~,
\end{equation}
where we have taken average values of the quark masses from
\cite{pdg}. Solving the system (\ref{U0}) and (\ref{mass}) gives
\begin{equation}
\label{AL} f_0 \approx 4.2 \times 10^{-3} ~, ~~~~~ |A_1| \approx
0.20244 /\varepsilon~,~~~~~ |A_{-1}| \approx 0.42717/\varepsilon~.
\end{equation}
Note these values of $|A_{\pm 1}|$ are consistent with the
condition in (\ref{norm}) which requires $|A_{\pm 1}|, ~|B_{\pm
1}| < \frac{0.4744}{\varepsilon}$. The largest mass corresponds to
the $l=0$ quantum number. This can be understood from the point of
view that this state has a non-zero effective wavefunction near
the brane, $\theta=0$, and thus has a large overlap with the
scalar field (\ref{Phi-0}). (By effective wavefunction we mean the
combination of the wavefunctions from (\ref{3-Psi}) and the square
root of the determinant from (\ref{determinant}). In this way the
singular $\sin \theta $ term cancels out). The $d$ and $s$ quarks,
which correspond to the $l=+1,-1$ states, have effective
wavefunctions which are zero at $\theta =0$ and thus have a smaller
overlap with the scalar field.

For mixings between the different families, characterized by
different angular momentum $l$, the selection rule (\ref{p-l})
indicates that we must consider components of the matrix
(\ref{U}), which have a nonzero upper index $p$. There are three
independent components whose indices are given by
\begin{eqnarray} \label{mixing}
U_{1,0}^{1} &=& U_{0,1}^{-1} = f_1\varepsilon^2 D_1
\frac{(1+\sqrt{2}) \pi}{2} (A_1^* A_0 + B^*_1 B_0) =
f_1\varepsilon^2D_{-1}\frac{(1+\sqrt{2})\pi}{2}(A_0^*A_1+B^*_0B_1)~,\nonumber\\
U_{0,-1}^{1} &=& U_{-1,0}^{-1} = f_1\varepsilon^2 D_1
\frac{(1+\sqrt{2}) \pi}{2} (A_0^* A_{-1} + B^*_0 B_{-1}) =
f_1\varepsilon^2D_{-1}\frac{(1+\sqrt{2})\pi}{2}(A_{-1}^*A_0+B^*_{-1}B_0)~,\\
U_{1,-1}^{2} &=& U_{-1,1}^{-2} = f_2 \varepsilon ^2 D_2 \sqrt{2}
\pi (A_1 ^* A_{-1} + B_1 ^* B_{-1} ) = f_2 \varepsilon ^2 D_{-2}
\sqrt{2} \pi (A_{-1} ^* A_1 + B_{-1} ^* B_1)~.\nonumber
\end{eqnarray}

From \cite{pdg} one finds that the mixing between the first and
second generation is of order $0.1$ (i.e. $V_{us} \approx 0.224$),
between the second and third generation of order $0.01$ (i.e.
$V_{cb} \approx 0.04$), and between the first and third generation
of order $0.001$ (i.e. $V_{ub} \approx 0.0036$). We take this
"up-down" sectors mixing as representing generic inter-family
mixing, since in our model we have only one flavor in each family
(the "down" sector and thus only neutral currents). Then from our
previous association of generations (first, second, third) with
the internal quantum number $l$ ($+1, -1, 0$) we arrive at the
following connections for the mixings from (\ref{mixing})
\begin{equation} \label{mixing2}
| U_{1,-1} ^2 | \rightarrow V_{us} \simeq 0.1 ~,~~~~ |U_{0,-1}^1|
\rightarrow V_{cb} \simeq 0.01 ~,~~~~~ |U_{1,0} ^1 | \rightarrow
V_{ub} \simeq 0.001
\end{equation}
In terms of ratios we want to fix $A_l, B_l$ such that from
(\ref{mixing}) we get
\begin{equation} \label{0.1}
\frac{|U_{1,0}^1|}{|U_{0,-1}^1|} \simeq 0.1~, ~~~~~
\frac{|U_{0,-1}^1|}{|U_{1,-1}^2|} \simeq 0.1~.
\end{equation}

To simplify the analysis we assume that all $A_l , B_l$ are purely
real. Then from (\ref{norm}) using (\ref{AL}) we have
\begin{equation}
\label{BL} |B_1| \approx  0.42907/\varepsilon~, ~~~~~ |B_{-1}|
\approx 0.20641 /\varepsilon~.
\end{equation}
Also we take $B_0 = k A_0$ where $k$ is some real constant, i.e.
from (\ref{norm}) $B_0$ is determined once $A_0$ is given.

Applying all this to the first condition from (\ref{0.1}) we find
\begin{equation}
\frac{|U_{1,0}^1|}{|U_{0,-1}^1|} = \frac{0.20244 + 0.42907 k}{0.42717
+ 0.20641 k} = 0.1~.
\end{equation}
Solving for $k$ gives $k=-0.39107~$.
For this value of $k$ we find from (\ref{norm}) that
\begin{equation}
A_0 = 0.52544/\varepsilon~, ~~~~~ B_0 = - 0.20548 /\varepsilon~.
\end{equation}
Inserting all these real values for $A_l, B_l$ into the second
condition from (\ref{0.1}) we find that
\begin{equation}
\label{mix123} \frac{|U_{0,-1}^1 |}{|U_{1,-1}^2|} =
1.065~\frac{f_1}{f_2}~.
\end{equation}
It is clear that if we set $f_1 /f_2 \sim 0.1$ (by adjusting
$\kappa _1$ and $\kappa_2$) we reproduce the mixings between the
different generations as given by the rough estimate
(\ref{mixing2}).


\section{Summary and Conclusions}

We have given a higher dimensional model to address the fermion
generation puzzle. Three zero mass modes arise in an "apple"
geometry given by (\ref{MetricAnsatz}) and (\ref{Solution}).
Exactly three zero modes are obtained by adjusting the shape of
the internal $2$-dimensional space via $b$ giving a profuse angle
rather than the more common case of a deficit angle. We interpret
these three zero mass modes as a toy model for the three
generations of fermions. This is a toy model since we do not
reproduce the full flavor structure of the Standard Model
fermions. For example in this paper we took the three zero mass
modes as the down quarks, $d,s,b$ leaving out the up quarks and
leptons. The family number in this model was the quantum number
$l$ associated with angular momentum of fermions with respect to
the extra $2$-space.

To give masses and mixings one had to couple the zero mass modes
to a scalar field. Thus in this model the masses and mixings arose
from the same mechanism. We demonstrated that one could get a
realistic mass spectrum and mixings by taking our zero mass modes
to be the family of down quarks. That we are able to reproduce a
realistic masses and mixings is not surprising since there are
number of free parameters involved especially in terms of the
normalization constants, $\kappa_p, A_l, B_l$ for the higher
dimensional wavefunctions. The central point of this paper was not
so much to obtain a realistic masses and mixings (since in any
case the model does not contain complete set of particles of the
Standard Model) but rather to give a higher dimensional model for
the fermion generation puzzle.

In addition to the zero mass modes there will be massive KK modes
whose masses will be of the order $1 / \varepsilon$. Here, since
$1 / \varepsilon \simeq 1$ TeV these massive KK modes would lie
well above the three zero mass modes even after they are given
masses. In any of these higher dimensional models used to address
the generation problem the internal space must be of a small
enough size so that the massive KK modes are well separated from
the zero mass modes after they are given a mass.

\vspace{0.5 cm}

{\bf Acknowledgment:} D.S. was supported by a CSU Fresno CSM
Summer Professional development grant during the course of this
work.



\begin{thebibliography}{99}

\bibitem{horisontal} C. D. Froggatt and H. B. Nielsen,
                     Nucl. Phys., {\bf B 147} (1979) 277;
                     M. Leurer, Y. Nir and N. Seiberg,
                     Nucl. Phys., {\bf B 398} (1993) 319;
                     L. E. Ibanez, G. G. Ross,
                     Phys. Lett., {\bf B 332} (1994) 100;
                     R. Barbieri, L. J. Hall, S. Raby and A. Romanino,
                     Nucl. Phys., {\bf B 493} (1997) 3;
                     J. K. Elwood, N. Irges and P. Ramond,
                     Phys. Rev. Lett., {\bf 81} (1998) 5064;
                     H. Fritzsch and Z.-z. Xing,
                     Prog. Part. Nucl. Phys., {\bf 45} (2000) 1;
                     M. S. Berger and K. Siyeon ,
                     Phys. Rev., {\bf D 71} (2005) 036005.

\bibitem{brane}  K. Akama,
                 in \emph{Gauge Theory and Gravitation} (Nara, Japan, 1982), eds.
                 K. Kikkawa, N. Nakanishi, and H. Nariai, Lecture Notes in Physics,
                 vol. 176;
                 V. A. Rubakov and M. E. Shaposhnikov,
                 Phys. Lett., {\bf B 125} (1983) 136;
                 Phys. Lett., {\bf B 125} (1983) 139;
                 M. Visser,
                 Phys. Lett., {\bf B 159} (1985) 22;
                 G. W. Gibbons and D. L. Wiltshire,
                 Nucl. Phys., {\bf B 287} (1987) 717;
                 N. Arkani-Hamed, S. Dimopoulos and G. Dvali,
                 Phys. Lett., {\bf B 429} (1998) 263;
                 I. Antoniadis, N. Arkani-Hamed, S. Dimopoulos and G. Dvali,
                 Phys. Lett., {\bf B 436} (1998) 257;
                 M. Gogberashvili,
                 Int. J. Mod. Phys., {\bf D 11} (2002) 1635;
                 Int. J. Mod. Phys., {\bf D 11} (2002) 1639;
                 Europhys. Lett., {\bf 49} (2000) 396;
                 Mod. Phys. Lett., {\bf A 14} (1999) 2025;
                 L. Randall and R. Sundrum,
                 Phys. Rev. Lett., {\bf 83} (1999) 3370;
                 Phys. Rev. Lett., {\bf 83} (1999) 4690.

\bibitem{local} A. Chodos and E. Poppitz,
                Phys. Lett., {\bf B 471} (1999) 119;
                A. G. Cohen and D. B. Kaplan,
                Phys. Lett., {\bf B 470} (1999) 52;
                B. Bajc and G. Gabadadze,
                Phys. Lett., {\bf B 474} (2000) 282;
                A. Pomarol,
                Phys. Lett., {\bf B 486} (2000) 153;
                R. Gregory,
                Phys. Rev. Lett., {\bf 84} (2000) 2564;
                Z. Chacko and A. E. Nelson,
                Phys. Rev., {\bf D 62} (2000) 085006;
                I. Oda,
                Phys. Rev., {\bf D 62} (2000) 126009;
                T. Gherghetta and M. Shaposhnikov,
                Phys. Rev. Lett., {\bf 85} (2000) 240;
                P. Kanti, R. Madden and K. A.Olive,
                Phys. Rev., {\bf D 64} (2001) 044021;
                S. Randjbar-Daemi and M. Shaposhnikov,
                Nucl. Phys., {\bf B 645} (2002) 188.

\bibitem{incr} M. Gogberashvili and P. Midodashvili,
               Phys. Lett., {\bf B 515} (2001) 447;
               Europhys. Lett., {\bf 61} (2003) 308;
               M. Gogberashvili and D. Singleton,
               Phys. Rev., {\bf D 69} (2004) 026004;
               P. Midodashvili,
               Europhys. Lett., {\bf 66} (2004) 478;
               Europhys. Lett., {\bf 69} (2005) 346;
               P. Midodashvili and L. Midodashvili,
               Europhys. Lett., {\bf 65} (2004) 640;
               I. Oda,
               Phys. Lett., {\bf B 571} (2003) 235.

\bibitem{Go-Si} M. Gogberashvili and D. Singleton,
                Phys. Lett., {\bf B 582} (2004) 95.

\bibitem{extra-mixing} K. R. Dienes, E. Dudas and T. Gherghetta,
                       Nucl.Phys., {\bf B 537} (1993) 47;
                       S. A. Abel and  S. F. King,
                       Phys. Rev., {\bf D 59} (1999) 095010;
                       K. Yoshioka,
                       Mod. Phys. Lett., {\bf A 15} (2000) 29;
                       N. Arkani-Hamed, L. J. Hall, D. R. Smith and  N. Weiner,
                       Phys. Rev., {\bf D 61} (2000) 116003;
                       D. E. Kaplan and T. M. P. Tait,
                       JHEP, {\bf 0111} (2001) 051;
                       N. Arkani-Hamed, S. Dimopoulos, G. Dvali and J. March-Russell,
                       Phys. Rev., {\bf D 65} (2002) 024032;
                       J. I. Silva-Marcos,
                       JHEP, {\bf 0703} (2007) 113;
                       I. Gogoladze, C.-A. Lee, Y. Mimura and Q. Shafi,
                       Phys. Lett., {\bf B 649} (2007) 212;
                       Y. Aghababaie, {\it et al.},
                       JHEP, {\bf 0309} (2003) 037;
                       M. Peloso, L. Sorbo and G. Tasinato,
                       Phys. Rev., {\bf D 73} (2006) 104025;
                       S. L. Parameswaran, S. Randjbar-Daemi and A. Salvio,
                       Nucl. Phys., {\bf B 767} (2007) 54.

\bibitem{fermi} N. Arkani-Hamed and M. Schmaltz,
                Phys. Rev., {\bf D 61} (2000) 033005;
                A. Mirabelli and M. Schmaltz,
                Phys. Rev., {\bf D 61} (2000) 113011;
                J. Schwindt and C. Wetterich
                Phys. Lett., {\bf B 578} (2004) 409.

\bibitem{Nav} I. Navarro,
              Class. Quant. Grav., {\bf 20} (2003) 3603;
              E. Papantonopoulos, A. Papazoglou and V. Zamarias,
              JHEP, {\bf 0703} (2007) 002.

\bibitem{rugby} S. Carroll and M. Guica,
                hep-th/0302067;
                J. Vinet and J. Cline,
                Phys. Rev., {\bf D70} (2004) 083514;
                J. Garriga and M. Porrati,
                JHEP,  {\bf 0408} (2004) 028;
                B. Himmetoglu and M. Peloso,
                Nucl. Phys., {\bf B 773} (2007) 84;
                C. P. Burgess, C. de Rham, D. Hoover, D. Mason and A. J. Tolley,
                JCAP, {\bf 0702} (2007) 009.

\bibitem{Ca-Hi} R. Camporesi and A. Higuchi,
                J. Geom. Phys., {\bf 20} (1996) 1.

\bibitem{Abr} A. A. Abrikosov (jr),
              Int. J. Mod. Phys., {\bf A 17} (2002) 885.

\bibitem{neronov} A. Neronov,
                  Phys. Rev., {\bf D 65} (2002) 044004;
                  S. Aguilar and D. Singleton,
                  Phys. Rev., {\bf D 73} (2006) 085007.

\bibitem{pdg} W.-M. Yao, {\it et al.},
              J. Phys., {\bf G 33} (2006) 1.

\end{thebibliography}
\end{document}